\documentclass[referee]{aa} 

\input psfig.sty

\begin{document}

\thesaurus{08     
              (06.22.1;  
               06.04.1;  
               11.04.1; 
)}

   \title{Theoretical Models for Classical Cepheids: VII. 
Metallicity effects on the Cepheid distance scale}

   \author{F. Caputo\inst{1}, M. Marconi\inst{2}, I. Musella\inst{2}, P. Santolamazza\inst{2}}

   \offprints{M. Marconi, marcella@na.astro.it}

   \institute{Osservatorio Astronomico di Roma, Via di Frascati 33,
I-00040 Monteporzio Catone, Italy \and Osservatorio Astronomico di Capodimonte, Via Moiariello 16,
I-80131 Napoli, Italy \\
              email: caputo@coma.mporzio.astro.it, marcella@na.astro.it, ilaria@na.astro.it}

   \date{Received 20 March 2000; accepted ........}

\titlerunning{Metallicity effects on the Cepheid distance scale}
\authorrunning{Caputo et al. }
\maketitle

   \begin{abstract}

We use theoretical Period-Luminosity and Period-Luminosity-Color
relations in the $VI$ passbands, as based on nonlinear, nonlocal
and time-dependent convective pulsating models, to predict
the reddening and true distance modulus of distant    
Cepheids observed with the Hubble Space Telescope.
By relying on the pulsating models with metal content $Z$=0.008, 
we find that the theoretical predictions agree to the values 
obtained 
by the Extragalactic Distance Scale Key Project
on the basis of empirical Period-Luminosity relations referenced to 
LMC variables. In the meantime, from the theoretical relations  
with $Z$=0.004 and 0.02 we find 
that the predicted 
$E(B-V)$ and $\mu_0$ decrease as the adopted metal content increases. 
This  
suggests a metallicity correction to LMC-based distances as given by 
$\Delta \mu_0$/$\Delta \log Z\sim-$0.27 mag dex$^{-1}$, where 
$\Delta \log Z$ is the difference between the metallicity of the 
Cepheids whose distance we are estimating  
and the LMC value $Z$=0.008. 
Such a theoretical correction appears supported by
an existing, although weak, correlation between the Cepheid distance 
and the [O/H] metallicity of   
galaxies within a given group or cluster, as well
as by a similar correlation between the $H_0$ estimate and 
the [O/H] metallicity 
of the galaxies which calibrate the SNIa luminosity. 
On the contrary, the metallicity correction 
earlier suggested on empirical grounds 
seem to be excluded.   
Eventually we suggest that the average value  
$<H_0>\sim$ 67 km s$^{-1}$ Mpc$^{-1}$ provided by the 
Key Project team should increase  {\it at least} up to 
$\sim$  69 km s$^{-1}$ Mpc$^{-1}$. 
Further 
observational evidences in support of 
the predicted scenario are finally presented.

   \end{abstract}

\newpage

\section{Introduction}

The Cepheid Period-Luminosity (PL) relation   
is a classical tool widely used to 
estimate the distance
to Local Group galaxies and to external galaxies with 
Hubble Space Telescope (HST) observations, 
as well as, through the calibration of secondary 
distance indicators, to even more distant stellar systems. 
This explains the 
dominant role of these variables on the route to 
solve cosmological problems 
such as the evaluation of the Hubble constant
$H_0$ and, in turn, of the age of the Universe.

The basic physics underlying the Cepheid
variability suggests that the pulsation period $P$ depends on the
mass, luminosity and effective temperature of the pulsator.
From stellar evolution theory one expects 
a close relation between mass and luminosity, and the natural result of
these theoretical prescriptions is 
a Period-Luminosity-Color (PLC)
relation, where the pulsator absolute magnitude
$M_j$ in each given photometric passband $j$ is a linear 
function of the period  and color index [$CI$], as
given by 

$$M_j=\alpha+\beta\log P+\gamma [CI]\eqno(1)$$ 

\noindent
However,
since the pulsation occurs within a finite
zone of the HR diagram, the color term in the PLC relation
is often neglected and Cepheid distances are usually estimated 
from the most favourite 
 PL relation

$$\overline{M_j}=a+b\log P,\eqno(2)$$ 

\noindent
where $\overline{M_j}$ is now the average of the Cepheid magnitudes
for each given period. It should be noticed that
whereas the PLC
relation holds for any individual pulsator, PL is a
``statistical'' solution which depends both on the pulsation boundaries
and on the distribution of pulsators within the instability
strip. This explains why 
distance determinations based on PL relations 
require statistically significant numbers of Cepheids in order to
reduce the effects of deviations from the ridge line.

The PL relation in bolometric magnitude is traditionally 
assumed to be metal-insensitive 
(see, e.g., Iben \& Renzini 1984, Freedman \& Madore 1990) 
and Cepheid distances 
are generally derived by adopting   {\it universal} 
PL relations at different wavelengths, having the 
slope provided by the Cepheids in 
the LMC and the zero-point referenced to
the LMC distance, as obtained with independent methods (see Freedman 1988;
Kennicutt et al. 1998; Walker 1999, and reference therein),  
or to calibrating
Galactic Cepheids (see Feast  \& Catchpole
1997; Lanoix et al. 1999). 

In the last years, we have 
deeply investigated the Cepheid pulsational behavior 
through the computations of  nonlinear, nonlocal and time-dependent 
convective pulsating models which take into account the coupling between 
pulsation and convection. 
With respect to    
linear-nonadiabatic models computed by different authors 
(e.g., Chiosi et al. 1993; Saio \& 
Gautschy 1998; Alibert et al. 1999), 
our theoretical approach allows reliable  
predictions on the temperature of 
both the blue and red edges of the instability strip, 
as well as about the amplitude and morphology 
of the light-curve (see Bono et al. 1999a [Paper I], 2000a [Paper III], 
2000b [Paper VI]). 
Concerning the assumptions on the input physics, computing  
procedures and the adopted mass-luminosity relation, the reader is referred to 
Paper I and Paper III, which  
contain also the detailed comparison of our models with the 
linear results in the literature.
Here we only 
remark that from  
pulsating models computed 
with three different chemical abundances ($Z$=0.004, 0.008 and 0.02) 
we derived that  
the {\it bolometric} magnitude of metal-rich variables
is, on average, fainter than that of metal-poor stars with the same period 
(Bono et al. 1999b [Paper II]). Moreover,   
we found that both the slope and zero-point of 
synthetic PL relations at 
different wavelengths depend on the pulsator metallicity, with the 
amplitude of the metallicity effect decreasing from visual to near-infrared 
magnitudes (see Paper II;  
Caputo et al. 2000 [Paper V]). Also the predicted 
PLC relations at the different wavelengths turned out to 
be, in various degrees, 
metallicity dependent: as an example, for a given period and 
$B-V$ color,     
metal-rich pulsators are brighter than metal-poor ones, 
whereas they are fainter if the $V-K$ color is adopted (see 
Paper V). On these grounds, 
it has been shown that 
Cepheid observations in three filters ($BVI$ or 
$BVK$) allow to simultaneously constrain 
distance, reddening and metallicity of 
variables in the Magellanic Clouds (Caputo et al.  
1999 [Paper IV]) and in the Milky Way 
(Caputo et al., in preparation). 

In this paper we will apply our theoretical 
scenario to the Cepheids  
observed within two HST surveys:
the ``Extragalactic Distance Scale Key Project'' (hereafter 
KP, see Freedman et 
al. 1994a) and the ``Type Ia Supernova 
Calibration'' (hereafter SNP, see Saha et al. 1994). 
The adopted 
procedure is 
briefly presented in Sect. 2 and the predicted 
reddenings and distances are 
given in Sect. 3. The metallicity 
effects on the Cepheid distance scale and on the value of the  
Hubble constant $H_0$ are discussed in 
Sect. 4. The concluding remarks close the paper. 

\section{The procedure} 

The results presented in Paper V predict that the intrinsic 
dispersion of the synthetic PL relations with fixed  
metallicity decreases 
with increasing the filter wavelength, in close agreement with the 
observed trend (see Madore \& Freedman 1991; Tanvir 1999). Moreover, 
we found that the predicted PLC relations 
are almost independent of the pulsator distribution within the 
instability strip and reproduce the tight 
correlation between pulsation period and photometric parameters 
of individual pulsators, as 
earlier suggested by Sandage (1958), Sandage \& Gratton (1963) and 
Sandage \& Tammann (1968). 

Since HST observations are in the two passbands $V$ and $I$,   
we adopt the 
predicted PL($I$) and PLC($VI$) relations given in 
Paper V and reported here in Table 1 and 
Table 2, respectively,    
to get the average reddening and true distance 
modulus of the Cepheid samples 
from the observed intensity-weigthed mean magnitudes 
$<V>$ and $<I>$, for each assumption about the Cepheid metallicity.  
With respect to the classical method based on PL($V$) and 
PL($I$), our procedure aims at reducing the 
spurious effects caused by both the large dispersion of  
the observed PL($V$) relation and 
the uncertainty on the predicted PL($V$) relation 
as due to different pulsator distributions (see 
Paper II and Paper V). 

The offset of each Cepheid from the predicted PL($I$) and 
PLC($VI$) relations gives 
the apparent distance moduli 
$\mu_1$=$\mu_0$+$A_I$ and   
$\mu_2$=$\mu_0$+$A_V-\gamma E(V-I)$, 
respectively. Adopting $A_V/E(B-V)$=3.1 and $A_I/E(B-V)$=1.88 
from Cardelli et al. (1989), 
the reddening $E(B-V)$ and true distance modulus $\mu_0$ of 
each Cepheid are
derived. Eventually, by averaging over the sample of variables, we  
provide the predicted mean values,   
for the three adopted metallicities.

\begin{table}[H]
\caption{Coefficients of theoretical PL relations in the $I$ band
for the labeled metal abundances.}

\vspace{0.5truecm}
\begin{tabular}{cccc}
\hline
\hline
Z & $a$ & $b$  & $\sigma$ \\
\hline
&&&\\   
\multicolumn{4}{c}{$\overline{M_I}$=$a$+$b$log$P$}\\
&&&\\
0.004&-2.00$\pm$0.01&-3.00$\pm$0.01&0.14\\
0.008&-2.03$\pm$0.01&-2.88$\pm$0.01&0.14\\
0.02&-2.18$\pm$0.01&-2.53$\pm$0.01&0.12\\
&&&\\
\hline
\hline
\end{tabular}\par
\end{table}

\begin{table}[H]
\caption{Coefficients of theoretical PLC relation in the $VI$ bands
for the labeled metal abundances.}
\begin{tabular}{ccccc}
\hline
\hline
Z & $\alpha$ & $\beta$  & $\gamma$ & $\sigma$\\
\hline
      &          &         &         &         \\
    \multicolumn{5}{c}{$<M_V>$=$\alpha$+$\beta$log$P$+$\gamma$[$<V>-<I>$]} \\
         &         &         &         &         \\
0.004 &-3.55 $\pm$0.03& -3.58 $\pm$0.03& 3.75 $\pm$0.07& 0.03 \\
0.008 &-3.54 $\pm$0.03& -3.59 $\pm$0.02& 3.74 $\pm$0.06& 0.03 \\
0.02  &-3.61 $\pm$0.03& -3.59 $\pm$0.04& 3.85 $\pm$0.09& 0.03 \\
         &         &         &         &         \\
\hline
\hline
\end{tabular}
\end{table}

\section{Predicted Cepheid distances to HST galaxies}

Table 3 lists the KP galaxies together with 
the color excess $E(B-V)$ and true distance modulus $\mu_0$, as 
given by the authors (see references in the last column), and the 
oxygen-to-hydrogen abundance ratios (O/H) of 
HII regions, as listed in Ferrarese et al. (1999).
For some of these galaxies the Cepheid data have 
been re-analyzed by the KP team and the new
reddening and distance values (Ferrarese et al. 1999) are given 
in italics in the same table. 
The galaxies marked with an asterisk, 
originally 
studied by the SNP group, were extracted from the HST 
archive and re-processed by the KP team (Gibson et al. 
2000). For them we 
list in Table 3 the KP values, while the SNP original distances  
are reported in Table 4. 

\begin{table}[H]
\caption{Empirical reddening and distance from KP studies. The values 
in italic deal with the revision of KP galaxies, as given in 
Ferrarese et al. (1999).
The asterisk marks the SNP galaxies re-processed by the KP 
team.}

\begin{tabular}{|c|c|c|c|c|}
\hline
\hline
&&&&\\[-15pt]
Galaxy & $E(B-V)$ & $\mu_0~~~~~$ &  12+log(O/H) & Reference \\[5pt]
\hline
\multicolumn{5}{c}{NGC 1023 Group} \\
NGC 925  & 0.13 $\pm$ 0.02 & 29.84   $\pm$0.16 & 8.55$\pm$0.15 & Silbermann et al. 1996  \\
\multicolumn{5}{c}{Fornax Cluster}\\
NGC 1326A  & 0.05 $\pm$ 0.09 &31.27 $\pm$0.21 & 8.50$\pm$0.15 & Prosser et al. 1999  \\[-6pt] 
 & {\it 0.00 $\pm$ 0.01} & {\it 31.43 $\pm$ 0.17 } &  & Ferrarese et al. 1999 \\[-4pt]
NGC 1365   & 0.12 $\pm$0.06 & 31.31 $\pm$0.20 & 8.96$\pm$0.20 & Silbermann et al. 1999 \\[-6pt]
 & {\it 0.15 $\pm$ 0.03} & {\it 31.39 $\pm$ 0.19} &  & Ferrarese et al. 1999 \\[-4pt]
NGC 1425   & 0.07 $\pm$0.03 & 31.73 $\pm$0.23 & 9.00$\pm$0.15 & Mould et al. 2000  \\[-6pt]
 & {\it 0.07 $\pm$ 0.03} & {\it 31.81 $\pm$ 0.17} &  & Ferrarese et al. 1999 \\
\multicolumn{5}{c}{M81 Group}\\
NGC 3031   & 0.03 $\pm$0.05 & 27.80 $\pm$0.20 & 8.75$\pm$0.15 & Freedman et al. 1994a \\
\multicolumn{5}{c}{NGC 3184  Group}\\
NGC 3198   & 0.05 $\pm$0.03 & 30.80 $\pm$0.23 & 8.60$\pm$0.15 & Kelson et al. 1999  \\[-4pt]
NGC 3319   & 0.06 $\pm$0.08 & 30.78 $\pm$0.17 & 8.38$\pm$0.15 & Sakai et al. 1999 \\
\multicolumn{5}{c}{Leo I Group}\\
NGC 3351   & 0.12 $\pm$ & 30.01 $\pm$0.19 & 9.24$\pm$0.20 & Graham et al. 1997 \\[-4pt]       
NGC 3368   & 0.11 $\pm$ & 30.20 $\pm$0.19 & 9.20$\pm$0.20 & Gibson et al. 2000  \\[-4pt] 
NGC 3627*   & 0.11 $\pm$0.04 & 30.06 $\pm$0.23 & 9.25$\pm$0.15 & Gibson et al. 2000  \\
\multicolumn{5}{c}{Coma I Cloud}\\
NGC 4414   & 0.01 $\pm$0.05 & 31.41 $\pm$0.19 & 9.20$\pm$0.15 & Turner et al. 1998 \\
\multicolumn{5}{c}{Coma II Cloud}\\
NGC 4725   & 0.19 $\pm$0.03 & 30.50 $\pm$0.16 & 8.92$\pm$0.15 & Gibson et al. 1999 \\[-6pt]
 & {\it 0.19 $\pm$ 0.03} & {\it 30.57 $\pm$ 0.18} &  & Ferrarese et al. 1999 \\
\multicolumn{5}{c}{M87 sub-cluster in the Virgo cluster}\\
NGC 4321   & 0.10 $\pm$0.06 & 31.04 $\pm$0.17 & 9.13$\pm$0.20 & Freedman et al. 1994b \\[-4pt]
NGC 4548   & 0.09 $\pm$0.03 & 31.01 $\pm$0.28 & 9.34$\pm$0.15 & Graham et al. 1999  \\
\multicolumn{5}{c}{NGC 4472 sub-cluster in the Virgo cluster}\\
NGC 4496A*  & 0.03 $\pm$0.05 & 31.02 $\pm$0.17 & 8.77$\pm$0.15 & Gibson et al. 2000 \\[-4pt]
NGC 4535   & 0.08 $\pm$0.02 & 31.02 $\pm$0.26 & 9.20$\pm$0.15 & Macri et al. 1999 \\[-6pt]
 & {\it 0.10 $\pm$ 0.03} & {\it 31.10 $\pm$ 0.17} &  & Ferrarese et al. 1999 \\[-4pt]
NGC 4536*   & 0.08 $\pm$0.03 & 30.95 $\pm$0.17 & 8.85$\pm$0.15 & Gibson et al. 2000 \\ 
\multicolumn{5}{c}{NGC 4649 sub-cluster in the Virgo cluster}\\
NGC 4639*   & 0.04 $\pm$0.03 & 31.80 $\pm$0.18      & 9.00$\pm$0.15 & Gibson et al. 2000  \\ 
\multicolumn{5}{c}{NGC 5128 Group}\\
NGC 5253*   & 0.10 $\pm$ & 27.61     $\pm$0.19 & 8.15$\pm$0.15 & Gibson et al. 2000 \\
\multicolumn{5}{c}{M101 Group}\\
NGC 5457 & 0.00 $\pm$0.04 & 29.34 $\pm$0.17 & 8.50$\pm$0.15 & Kelson et al. 1996\\
\multicolumn{5}{c}{NGC 7331 Group}\\
NGC 7331   & 0.15 $\pm$0.05 & 30.89     $\pm$0.14 & 8.67$\pm$0.15 & Hughes et al. 1998 \\
\multicolumn{5}{c}{Other}\\
IC 4182*   & -0.02 $\pm$ 0.03 & 28.36     $\pm$0.18 & 8.40$\pm$0.20 & Gibson et al. 2000 \\[-4pt]
NGC 2090   & 0.07 $\pm$0.02 & 30.45 $\pm$0.16 & 8.80$\pm$0.15 & Phelps et al. 1998 \\ [-4pt]
NGC 2541   & 0.08 $\pm$0.05 & 30.47 $\pm$0.11 & 8.50$\pm$0.15 & Ferrarese et al. 1998 \\[-4pt]
NGC 3621   & 0.23 $\pm$0.03 & 29.13 $\pm$0.18 & 8.75$\pm$0.15 & Rawson et al. 1997  \\
\hline
\hline
\end{tabular}
\end{table}

\begin{table}[H]
\caption{Empirical Cepheid distance to SNP galaxies, as published 
in the original papers by the SNP team.}

\begin{tabular}{|c|c|c|}
\hline
\hline
&&\\[-15pt]
Galaxy & $\mu_0~~~~~$ & Reference  \\[5pt]
\hline
NGC 3627   & 30.22 $\pm$0.12  &  Saha et al. 1999  \\
NGC 4496A  & 31.03 $\pm$0.14  &  Saha et al. 1996b \\
NGC 4536   & 31.10 $\pm$0.13  &  Saha et al. 1996a \\
NGC 4639   & 32.03 $\pm$0.22  &  Saha et al. 1997 \\
NGC 5253   & 28.08 $\pm$0.10  &  Saha et al. 1995 \\
IC 4182    & 28.36 $\pm$0.09  &  Saha et al. 1994 \\ 
\hline
\hline
\end{tabular}
\end{table}

\begin{table}[H]

\caption{Predicted Cepheid reddening and distance of   
KP galaxies for the labelled metal contents. The
distance moduli based on the selected data used by Ferrarese et al. (1999) 
in their 
revision studies are labeled in italics.
For the galaxies marked with the asterisk, originally 
observed by the SNP team, the 
photometric data provided by the KP group are adopted.}

\vspace{0.5truecm}

\begin{tabular}{|l|rr|rr|rr|}
\hline
\hline
&&&&&\\[-15pt]   
&\multicolumn{2}{c|}{Z=0.004}&\multicolumn{2}{c|}{Z=0.008}&\multicolumn{2}{c|}{Z=0.02}\\[5pt]
\hline
Galaxy    &  $E(B-V)$  & $\mu_0~~~~~$  &  $E(B-V)$  & $\mu_0~~~~~$ &  $E(B-V)$  & $\mu_0~~~~~$\\[5pt]
\hline
&&&&&&\\[-15pt]     
  IC 4182* & -0.06 $\pm$ 0.05 & 28.51 $\pm$ 0.18 &-0.09 $\pm$ 0.05& 28.48 $\pm$ 0.18 & -0.15 $\pm$ 0.04 & 28.37 $\pm$ 0.15 \\
  NGC 925  &  0.10 $\pm$ 0.05 & 29.98 $\pm$ 0.17 & 0.06 $\pm$ 0.05& 29.94 $\pm$ 0.17 & -0.02 $\pm$ 0.04 & 29.80 $\pm$ 0.15 \\
  NGC 1326A&  0.02 $\pm$ 0.05 & 31.42 $\pm$ 0.18 &-0.03 $\pm$ 0.05& 31.38 $\pm$ 0.18 & -0.11 $\pm$ 0.04 & 31.22 $\pm$ 0.16 \\ 
         &  & {\it 31.58 $\pm$ 0.18} & & {\it 31.53 $\pm$ 0.18} & & {\it 31.35 $\pm$ 0.16} \\
  NGC 1365 &  0.07 $\pm$ 0.05 & 31.49 $\pm$ 0.17 & 0.02 $\pm$ 0.05& 31.45 $\pm$ 0.17 & -0.07 $\pm$ 0.04 & 31.27 $\pm$ 0.15 \\
         &  & {\it 31.50 $\pm$ 0.18} & & {\it 31.45 $\pm$ 0.18} & & {\it 31.27 $\pm$ 0.16} \\
  NGC 1425 &  0.04 $\pm$ 0.05 & 31.85 $\pm$ 0.18 & 0.00 $\pm$ 0.05& 31.80 $\pm$ 0.17 & -0.10 $\pm$ 0.04 & 31.63 $\pm$ 0.15 \\
         &  & {\it 31.88 $\pm$ 0.18} & & {\it 31.83 $\pm$ 0.18} & & {\it 31.64 $\pm$ 0.15} \\
  NGC 2090 &  0.06 $\pm$ 0.05 & 30.53 $\pm$ 0.17 & 0.01 $\pm$ 0.05& 30.49 $\pm$ 0.17 & -0.07 $\pm$ 0.04 & 30.34 $\pm$ 0.15 \\
  NGC 2541 &  0.07 $\pm$ 0.05 & 30.55 $\pm$ 0.18 & 0.02 $\pm$ 0.05& 30.51 $\pm$ 0.18 & -0.07 $\pm$ 0.04 & 30.35 $\pm$ 0.16 \\
  NGC 3198 &  0.04 $\pm$ 0.05 & 30.95 $\pm$ 0.18 & -0.01 $\pm$0.05& 30.90 $\pm$ 0.18 &  -0.10$\pm$ 0.05 & 30.75 $\pm$ 0.16 \\
  NGC 3031  &  0.07 $\pm$ 0.05 & 27.89 $\pm$ 0.19 & 0.03 $\pm$ 0.05 & 27.85 $\pm$ 0.19 & -0.05 $\pm$ 0.05 & 27.70 $\pm$ 0.16 \\
  NGC 3319 &  0.02  $\pm$0.05 &30.94  $\pm$ 0.19& -0.02$\pm$0.05 &30.89  $\pm$0.19  & -0.11 $\pm$ 0.05 &30.74 $\pm$0.17  \\
  NGC 3351 &  0.09 $\pm$ 0.06 & 30.18 $\pm$ 0.19 & 0.05 $\pm$ 0.06 & 30.14 $\pm$ 0.19 & -0.02 $\pm$ 0.05 & 30.00 $\pm$ 0.17 \\
  NGC 3368   &  0.09 $\pm$ 0.05 & 30.17 $\pm$ 0.20 & 0.04 $\pm$ 0.05 & 30.13 $\pm$ 0.19 & -0.04 $\pm$ 0.05 & 29.97 $\pm$ 0.17\\
  NGC 3621&  0.20 $\pm$ 0.05 & 29.31 $\pm$ 0.18 & 0.15 $\pm$ 0.05 & 29.27 $\pm$ 0.18 &  0.07 $\pm$ 0.04 & 29.12 $\pm$ 0.16 \\
  NGC 3627* & 0.12 $\pm$ 0.05 & 29.93 $\pm$ 0.18 & 0.08 $\pm$ 0.05 & 29.88 $\pm$ 0.18 &  -0.01 $\pm$ 0.04 & 29.72 $\pm$ 0.16  \\
  NGC 4321  &  0.07 $\pm$ 0.05 & 31.04 $\pm$ 0.18 & 0.02 $\pm$ 0.05 & 30.99 $\pm$ 0.18 & -0.07 $\pm$ 0.04 & 30.82 $\pm$ 0.16\\
  NGC 4414  &  0.04 $\pm$ 0.07 & 31.42 $\pm$ 0.21 &-0.01 $\pm$ 0.07 & 31.36 $\pm$ 0.21 & -0.11 $\pm$ 0.06 & 31.18 $\pm$ 0.19 \\
  NGC 4496A*&  0.04 $\pm$ 0.04 & 31.04 $\pm$ 0.17 &-0.01 $\pm$ 0.04 & 30.99 $\pm$ 0.17 & -0.10 $\pm$ 0.04 & 30.83 $\pm$ 0.15 \\
  NGC 4535  &  0.06 $\pm$ 0.05 & 31.12 $\pm$ 0.17 & 0.02 $\pm$ 0.05 & 31.07 $\pm$ 0.17 & -0.08 $\pm$ 0.04 & 30.90 $\pm$ 0.15 \\
         &  & {\it 31.16 $\pm$ 0.18} & & {\it 31.11 $\pm$ 0.18} & & {\it 30.93 $\pm$ 0.16} \\
  NGC 4536* &  0.06 $\pm$ 0.05 & 31.01 $\pm$ 0.17 & 0.02 $\pm$ 0.05 & 30.96 $\pm$ 0.17 & -0.07 $\pm$ 0.04 & 30.79 $\pm$ 0.15 \\
  NGC 4548  &  0.07 $\pm$ 0.05 & 31.15 $\pm$ 0.18 & 0.02 $\pm$ 0.05 & 31.10 $\pm$ 0.18 & -0.06 $\pm$ 0.05 & 30.95 $\pm$ 0.16 \\
  NGC 4639* &  0.05 $\pm$ 0.05 & 31.83 $\pm$ 0.19 &-0.01 $\pm$ 0.05 & 31.78 $\pm$ 0.19 & -0.11 $\pm$ 0.05 & 31.59 $\pm$ 0.16 \\
  NGC 4725  &  0.15 $\pm$ 0.05 & 30.60 $\pm$ 0.18 & 0.10 $\pm$ 0.05 & 30.55 $\pm$ 0.18 &  0.01 $\pm$ 0.04 & 30.39 $\pm$ 0.16 \\
         &  & {\it 30.65 $\pm$ 0.18} & & {\it 30.60 $\pm$ 0.18} & & {\it 30.42 $\pm$ 0.16} \\
  NGC 5253* &  0.08 $\pm$ 0.06 & 27.75 $\pm$ 0.21 & 0.05 $\pm$ 0.06 & 27.72 $\pm$ 0.21 &  0.00 $\pm$ 0.06 & 27.62 $\pm$ 0.21 \\
  NGC 5457  &0.01 $\pm$ 0.05 & 29.45 $\pm$ 0.18 &-0.04 $\pm$ 0.05 & 29.40 $\pm$ 0.18 &-0.13 $\pm$ 0.05 & 29.23 $\pm$0.16\\
  NGC 7331 &  0.13 $\pm$ 0.06 & 31.01 $\pm$ 0.20 & 0.09 $\pm$ 0.06 & 30.97 $\pm$ 0.20 & -0.00 $\pm$ 0.06 & 30.81 $\pm$ 0.18 \\[5pt]
\hline
\hline
\end{tabular}
\end{table}


\begin{table}

\caption{Predicted Cepheid distance to SNP galaxies for   
the labelled metal contents. 
The full sets of SNP photometric data are adopted.
The results obtained from the selected data used by the SNP team 
are labeled in italics.}

\vspace{0.5truecm}

\begin{tabular}{|c|c|c|c|}
\hline
\hline
&&&\\[-15pt]
&$Z=0.004$&$Z=0.008$&$Z=0.02$\\[5pt]
\hline
Galaxy    &  $\mu_0$ & $\mu_0$ & $\mu_0$\\[5pt]
\hline
&&&\\[-15pt]
   NGC 3627 &  29.94 $\pm$ 0.19 & 29.89 $\pm$ 
 0.19 & 29.73 $\pm$ 0.17\\
 & {\it 30.37 $\pm$ 0.18} & {\it 30.33 $\pm$ 0.18} & {\it 30.16 $\pm$ 0.16} \\
   NGC 4496A &  31.07 $\pm$ 0.17 & 31.02 $\pm$
 0.17 & 30.86 $\pm$ 0.15 \\
 & {\it 31.12 $\pm$ 0.17} & {\it 31.07 $\pm$ 0.17} & {\it 30.89 $\pm$ 0.15} \\
  NGC 4536 & 30.75 $\pm$ 0.18 & 30.71 $\pm$
 0.18 & 30.55 $\pm$ 0.16 \\
 & {\it 30.93 $\pm$ 0.20} & {\it 30.88 $\pm$ 0.20} & {\it 30.70 $\pm$ 0.18} \\
   NGC 4639 & 31.85 $\pm$ 0.23 & 31.80 $\pm$
 0.23 & 31.63 $\pm$ 0.21 \\
 & {\it 31.94 $\pm$ 0.21} & {\it 31.89 $\pm$ 0.21} & {\it 31.71 $\pm$ 0.19} \\
   NGC 5253 & 27.81 $\pm$ 0.24 & 27.78 $\pm$
 0.25 &  27.70 $\pm$ 0.24 \\
 & {\it 27.99 $\pm$ 0.22} & {\it 27.96 $\pm$ 0.22} & {\it 27.89 $\pm$ 0.21} \\
  IC 4182 & 28.81 $\pm$ 0.19 & 28.77 $\pm$
 0.19 & 28.66 $\pm$ 0.17 \\
 & {\it 28.70 $\pm$ 0.19} & {\it 28.66 $\pm$ 0.19} & {\it 28.53 $\pm$ 0.17} \\[5pt]
\hline
\hline
\end{tabular}
\end{table}

\begin{figure}
\vspace{-2truecm}
\psfig{figure=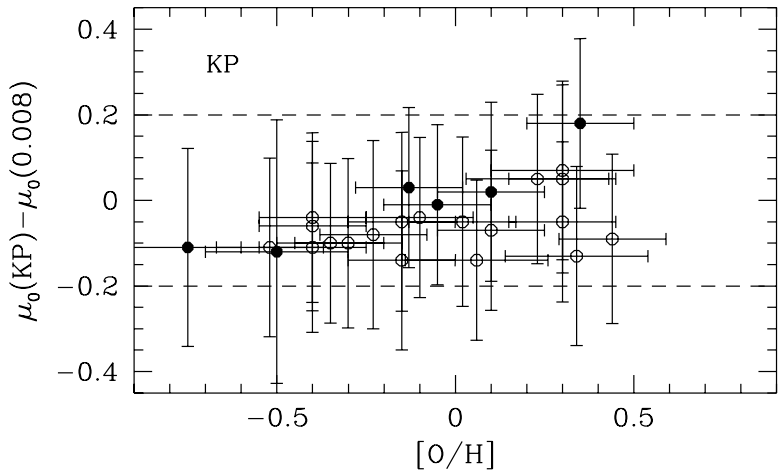,width=14truecm,angle=0}
\begin{flushleft}
\end{flushleft}

\caption{The difference between the empirical Cepheid distance by 
KP studies and 
the predicted value with $Z=0.008$, plotted against the O/H metallicity 
of the host galaxy. The Cepheid samples are original KP data (open 
circles) or refer to 
SNP galaxies extracted from the HST archive and 
re-processed by the KP team (filled circles). 
The two dashed lines depict a 
discrepancy on the distance of the order of $\pm$10\%.}

\end{figure}

\begin{figure}
\vspace{-2truecm}
\psfig{figure=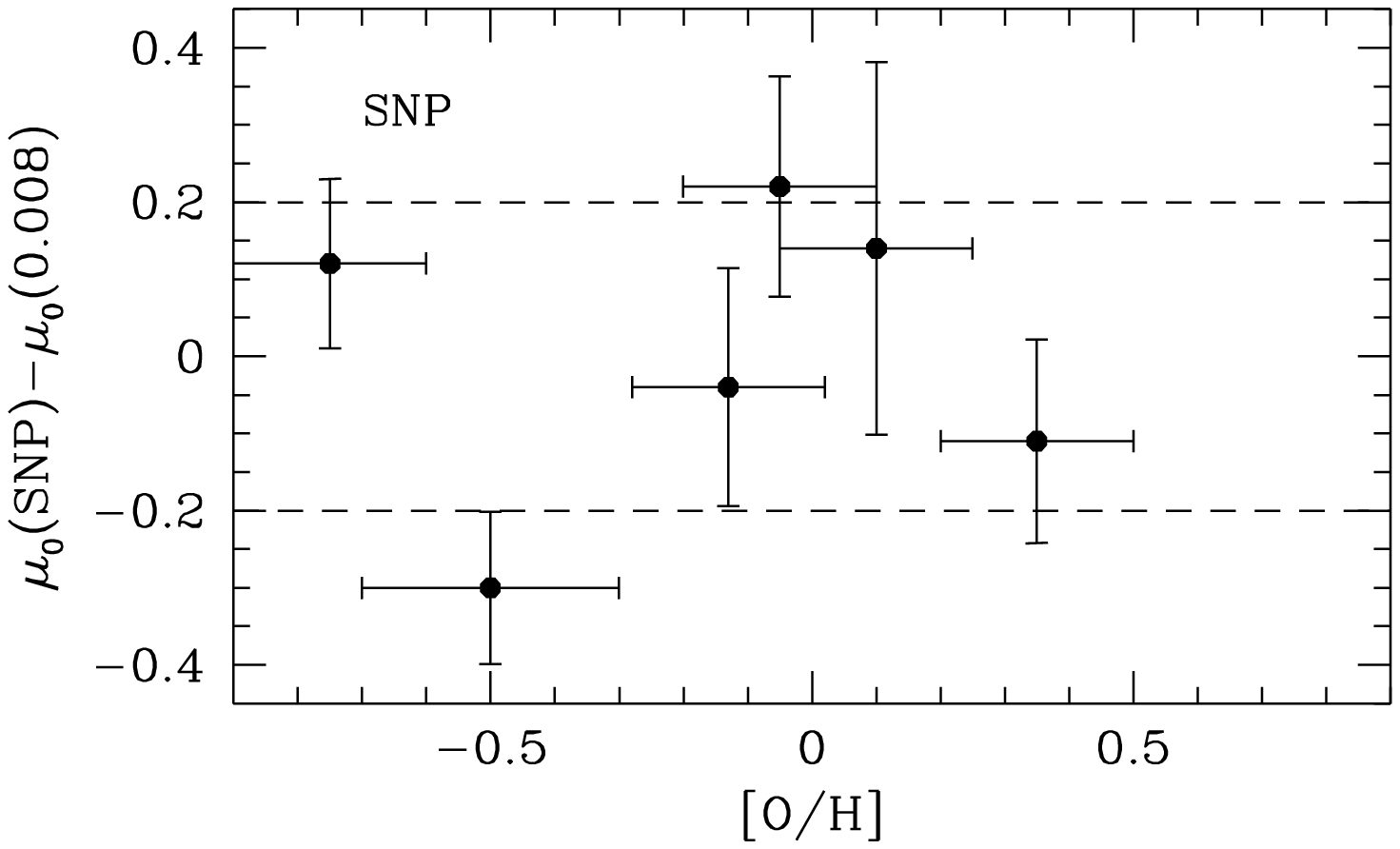,width=14truecm,angle=0}
\begin{flushleft}
\end{flushleft}

\caption{As in Fig. 1, but with the Cepheid distances by 
the SNP group.}

\end{figure}

By applying the theoretical relations in Table 1 and Table 2 to 
the full samples of Cepheids provided by KP and SNP studies  
(references in Table 3 and Table 4), 
we derive the predicted reddenings and true distance 
moduli listed in Table 5 and Table 
6, respectively, for each of the three selected 
metal abundances. To avoid any bias, the 
predicted true distance modulus includes the correction for 
negative reddening, if present. The total  
errors to the predicted values are due to the 
observed scatter in the PL and PLC 
fits added in quadrature to the 
intrinsic dispersion of the theoretical relations. 

Considering that both KP and SNP distances adopt PL($V$) 
and PL($I$)  relations based 
on the Cepheids in the LMC, and since the average metallicity 
of LMC Cepheids is $<Z>\sim$0.008  
(Luck et al. 1998), let us first compare   
the predicted reddening and distance 
with $Z$=0.008 to the LMC-based empirical 
values. It seems worth mentioning that we have already found a 
fair agreement between the empirical PL($V$) and PL($I$) relations 
and our predictions with $Z$=0.008 (see Paper V).

As for the galaxies in Table 3, we obtain that empirical and  
predicted results with $Z$=0.008 (Table 5)    
are in mutual agreement, 
with a difference (empirical minus 
predicted) of $\Delta \mu_0\sim-$0.05 mag and 
$\Delta E(B-V)\sim+$0.05 mag, in the mean. 
We show in Fig. 1 that the discrepancy  
is for each galaxy well within a total   
uncertainty on the 
distance of the order of 10\% (dashed lines). This is a reassuring 
result as to the reliability of the adopted method of 
analysis and our pulsating models with $Z$=0.008. 
  
 Moreover, 
one has to consider that 
both KP and SNP studies assume  
$\mu_{0,LMC}$=18.50 mag and $E(B-V)_{LMC}$=0.10 mag, 
and that the above differences
of $\Delta \mu_0\sim-$0.05 mag and $\Delta E(B-V)\sim+$0.05 mag 
with respect to the predictions with $Z$=0.008 
could be easily removed by slightly increasing (decreasing)  
the adopted LMC distance (reddening). 
On the other hand, the theoretical results depend on 
assumptions such as the adopted mass-luminosity 
ratio, bolometric corrections and temperature-color 
relations. Thus, revisions of input physics 
and/or atmosphere models might modify the results in 
Table 5 and Table 6, but with marginal effects on  
the relative 
distances and reddenings. As a fact, by applying our method 
to the LMC Cepheids (Madore \& Freedman 1991) we get 
$\mu_{0,LMC}\sim$18.6 mag and $E(B-V)_{LMC}\sim$0.03 mag. 
Accounting for these results, the actual discrepancy between 
empirical and predicted results is
$\Delta \mu_0\sim+$0.05 mag and
$\Delta E(B-V)\sim-$0.02 mag, in the mean.
   
On the contrary, it seems important to notice that different distributions 
of Cepheids within the instability strip may produce non-systematic 
effects on both distance and reddening. As a matter of example, 
let us consider a Cepheid with absolute magnitudes $M^*_V$ and $M^*_I$, 
true distance modulus $\mu^*_0$ and reddening $E^*(B-V)$, and 
let us assume that  $M^*_j$=$\overline{M_{j}}-\epsilon_{j}$, 
where $\overline{M_{j}}$ is given by the PL($j$) relation. From the   
PL($V$)+PL($I$) procedure one derives   
$E(V-I)=E^*(V-I)-(\epsilon_{V}-\epsilon_{I})$, while 
following our PL($I$)+PLC($VI$) method one has 
$E(V-I)=E^*(V-I)-0.36\epsilon_{I}$. Since  
Cepheids close to the blue edge of the instability 
strip have positive values of $\epsilon_{j}$, with   
$\epsilon_{V}>\epsilon_{I}$  
as a consequence of the fact 
that the width of the instability strip decreases 
towards longer wavelengths (see Paper V), the result 
is that the actual reddening of Cepheids near the pulsation blue 
edge is underestimated, whereas the opposite occurs for variables 
close the red edge. 
Concerning the distance,
from the PL($V$)+PL($I$) procedure one has
$\mu_0=\mu^*_0+1.54\epsilon_{V}-2.54\epsilon_{I}$,
while the PL($I$)+PLC($VI$) method yields
$\mu_0=\mu^*_0-0.45\epsilon_I$.   
With $\epsilon_{V}\sim$ 0.35 mag and 
$\epsilon_{I}\sim$ 0.25 mag for Cepheids close to 
the pulsation blue edge\footnote{Average values with log$P\sim$
1.0-1.5.}, 
the derived reddening and true distance 
modulus are $\sim$ 0.10 mag lower than the actual values, whatever 
is the adopted method of analysis. 
On these grounds,    
one understands that statistically significant samples of 
variables are required 
in order to get accurate distance and reddening 
determinations. Otherwise, selection effects on 
distant Cepheids (the variables close to the blue edge are brighter than 
those near the red edge, for fixed period) lead to underestimate 
both the distance and reddening. 
   
As already  
mentioned, for some of the KP galaxies in Table 3 
the Cepheid data have been reviewed by the same team,  
leading to a slight systematic increase on the distance. 
From the data listed in italics in Table 3 one 
finds that the average upward revision is of the order of $\sim$0.09 mag. 
As discussed in Ferrarese et al. (1999), this is a
consequence of the adopted lower cut-off in period, as 
the empirical distance modulus increases when the lower period cut-off
is moved from 10 to 20-25 days. On the theoretical side,
we show in Table 5 (see values in italics)  that using the 
same selection  
as in Ferrarese et al. (1999) causes a systematic 
increase of the predicted true distance modulus by only $\sim$0.05 mag. 
This suggests that the
cut-off in period has minor influence upon our PL($I$)+PLC($VI$) 
procedure, supporting the belief 
that robust distance determinations should adopt only  
PLC relations, for which Cepheid observations in at
least three filters are needed.

Concerning the SNP galaxies, we give in Table 6 the predicted 
distances as obtained from the full set of Cepheid data 
provided in the original papers, while the values in 
italics 
refer to the Cepheid samples as selected by the SNP team. 
One notices  
that the adopted selection criteria 
yield rather discordant effects on the predicted distance. With 
$Z$=0.008, the 
true distance modulus of four galaxies increases 
(from 0.05 mag for NGC 4496A to 0.44 mag for NGC 3627), whereas 
it decreases by $-$0.11 mag for IC 4182. Moreover, 
limiting the comparison to the SNP empirical distances obtained  
from 
the selected samples (Table 4), the difference with respect to    
the predicted distance modulus with $Z$=0.008 (in 
italics in Table 7) 
ranges from $-$0.30 mag (IC 4182) 
to +0.22 mag (NGC 4536), which is somehow larger than that 
(from $-$0.12 mag for IC 4182 to +0.18 mag for NGC 3627) found 
by adopting KP results for the same galaxies (see Table 3 and Table 5). 

It is known that the disagreement     
between SNP and KP distances to the same galaxy 
(up to 0.47 mag for NGC 5253) deals with 
different pipeline processing procedures,  
selection criteria and data reduction strategies, and the 
interested reader can find
a detailed discussion
in Gibson et al. (2000) and Tammann et al. (2000). Here we 
have to face up to the evidence that by using the Cepheid data 
selected by SNP studies the overall discrepancy between 
empirical and predicted distances with $Z$=0.008 is somehow 
larger (see Fig. 2) than that found with KP 
data  
(filled circles in 
Fig. 1). For this reason, the rest of our analysis will 
deal exclusively with the results of KP studies.     

Throughout the above discussion we used the Cepheid predicted distance 
with $Z$=0.008, neglecting the evidence  
that the oxygen abundance 
of the galaxies in Table 3  varies from [O/H]$\sim-$0.50 to 
$\sim$+0.40\footnote{[O/H]=log(O/H)-log(O/H)$_{\odot}$, 
with log(O/H)$_{\odot}$=-3.10.}. Since for the LMC  
[O/H]$\sim-$0.40 (Pagel et al. 1978), 
one has that the HII metallicity of the Cepheid 
fields observed with HST may be up to 0.8 dex higher than 
the ``reference'' sample of LMC variables. 
 
On the other hand, the data in Table 5 show that the predicted 
Cepheid reddening and true distance modulus decrease 
as the adopted metallicity $Z$ 
increases. This is a consequence of the fact that our metal-rich 
pulsating models are, on average, fainter than metal-poor ones. Resuming the 
discussion in Sect. 3, let us assume that for our Cepheid holds   
$\epsilon_V=\epsilon_I=$0 with respect to the PL relations with 
$Z$=0.008. This yields $E(V-I)_{0.008}$=$E^*(V-I)$ and 
$\mu_{0,0.008}$=$\mu^*_0$. However, the variable is brighter with 
respect to the PL relations with $Z$=0.02, by  
$\epsilon_V\sim$0.25 mag and $\epsilon_I\sim$0.20 at log$P\sim$1.0 
(see Paper V). As a consequence, $E(V-I)_{0.02}\sim E^*(V-I)-$0.06 mag and 
$\mu_{0,0.02}\sim\mu^*_0-$0.11 mag. 
The decrease of the 
reddening with increasing 
the adopted metallicity yields that almost all the galaxies show 
unpleasant  
negative values of $E(B-V)$ with $Z$=0.02. However, 
as already mentioned, 
our theoretical scenario suggests $E(B-V)\sim$0.03 mag for the 
Cepheids in the LMC, while both the KP and SNP studies assume 
$E(B-V)_{LMC}$=0.10 mag. Thus, were our predictions calibrated on 
this galaxy, then all the reddening values in Table 5 and Table 6 
should be increased by 0.07 mag. 

Since our purpose is the 
evaluation of the metallicity effects on the distance scale,  
in the mean we 
derive $\delta \mu_0/\delta \log Z=-$0.27$\pm$0.04 mag dex$^{-1}$, at least 
within the explored range 0.004$\le Z \le$ 0.02. The sense of 
this result is that when universal relations are used,  
the Cepheids more metal-rich than the 
calibrators will appear spuriously {\it brighter}. On this basis, 
the predicted correction (in magnitude) 
to empirical LMC-calibrated distance moduli is 
given by   

$$c=-0.27\Delta \log Z,\eqno(3)$$ 

\noindent
where $\Delta \log Z$ is the difference 
between the Cepheid metallicity and the LMC value of $Z$=0.008. 
 
In the following section we investigate into the effects of this 
metallicity correction to KP distances by 
assuming that the Cepheid metal content scales with the HII region 
metallicity, i.e. adopting $\Delta \log Z$=$\Delta$[O/H], 
the difference between the galaxy [O/H]  
abundance and that of the LMC ([O/H]=$-$0.40). 
Observational 
pros and cons for this assumption will be discussed.

\begin{figure}
\psfig{figure=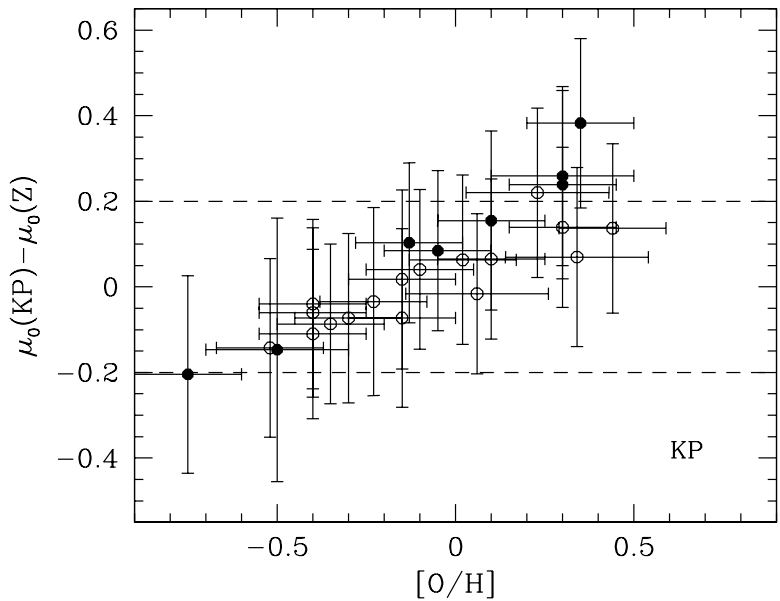,width=14truecm,angle=0}
\caption{The difference between KP empirical distances and
predicted metallicity-corrected values, plotted
against the HII metallicity
of the host galaxy. The filled circles depict the eight galaxies
which give Cepheid-calibrated SNIa luminosities.}
\end{figure}

\section{Metallicity effects on Cepheid distances and $H_0$}  

Figure 3 shows the difference between the empirical  
KP data\footnote{For sake of homogeneity, we adopt the Cepheid 
distances listed in Ferrarese et al. (1999).}  
and the predicted true distance modulus when 
the oxygen abundance of the Cepheid fields is taken into 
account. As expected, the  
almost flat distribution in Fig.1 now shows a 
clear correlation to the HII region metallicity, 
with the discrepancy between LMC-calibrated and theoretical distances 
increasing when moving from metal-poor to metal-rich galaxies. 
However, one notices that for very few galaxies the discrepancy  
overcomes the threshold of 10\% (dashed line). 
Of particular interest is the behaviour of the eight 
galaxies (the six SNP galaxies 
plus NGC 3368 and NGC 4414; filled circles) 
which give Cepheid-calibrated 
SNIa luminosities (see later). 

\begin{figure}
\psfig{figure=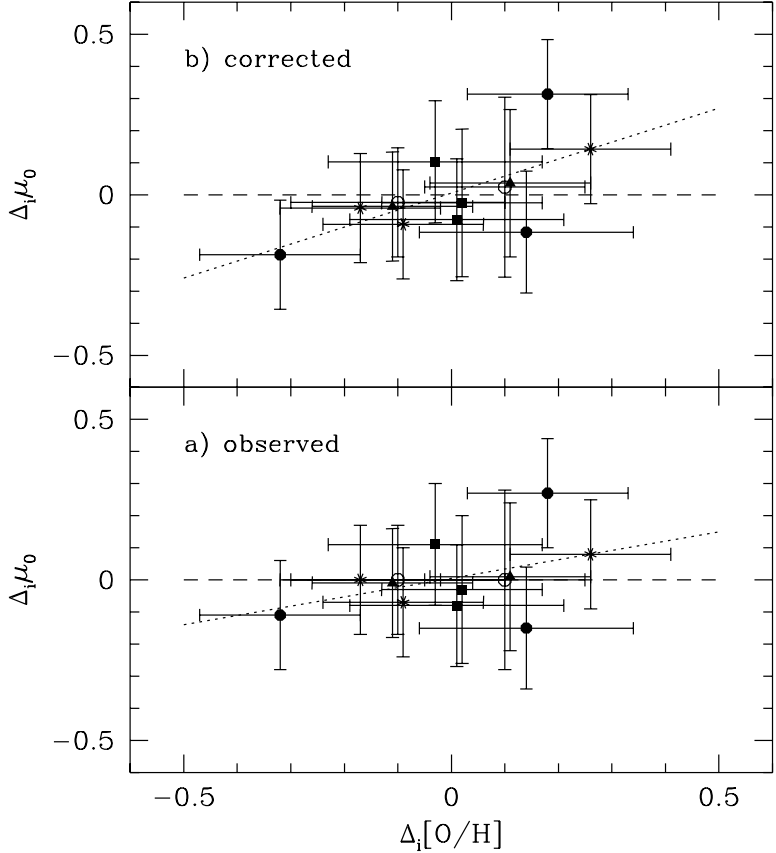,width=14truecm,angle=0} 
\caption[h]{a) Residuals of each galaxy from the average true distance 
modulus and O/H metallicity of the host cluster or group. 
Symbols as follows: Fornax cluster: filled circles, NGC 3184 
group: triangles, Leo I 
group: squares, M87 sub-cluster: open circles, 
NGC 4472 sub-cluster: asterisks. The dotted line is the best 
fit to the points [Eq. (4)].
b) As below, but with the Kennicutt et al. (1998) metallicity 
correction to the absolute distance moduli [Eq. (5)].}
\end{figure}

As a first straightforward test to the actual occurrence of a 
metallicity effect,  
we consider the KP distance to galaxies members of groups or clusters. 
The  lower panel of Fig. 4 shows the residuals 
$\Delta_i\mu_0$ and $\Delta_i$[O/H] of each galaxy from the 
distance modulus and O/H metallicity as averaged 
over the galaxies of the same group or cluster. A correlation between the  
metallicity and distance deviations from the mean values can 
be detected, best-fitted by the relation 
(dotted line) 

$$\Delta_i\mu_0=0.28\Delta_i [O/H],\eqno(4)$$

\noindent
in the sense that galaxies whose O/H metallicity is larger than the 
average appear to have a {\it larger} distance. We believe that 
depth-effects within a given group or cluster cannot be invoked 
since 
there is no reason for which the metal-richest galaxies are also the  
most distant ones. 

This unexpected result, which  
appears surprisingly in agreement with our predicted correction 
[Eq. (3)], disagrees with 
earlier observational clues. It is known that studies of 
different fields in M31 (Freedman \& Madore 1990) and M101 
(Kennicutt et al. 1998) suggested an opposite metallicity correction 
on distance (see also Kochanek 1997; Sasselov et al. 1997). 
Following Kennicutt et al. (1998), the correction (in 
magnitude) to the true distance modulus is given by 

$$c=+0.24\Delta [O/H]\eqno(5)$$ 

\noindent
We show in the upper panel of the same Fig. 4 that adopting 
such an empirical correction would imply an even stronger 
correlation between distance and metallicity, with a slope of +0.53 
(dotted line)
hard to accept.   

Applying the predicted metallicity 
correction to the absolute distance moduli [Eq. (3)], 
obviously results 
in a correction to the $H_0$-values based on LMC-calibrated
distances. Assuming 
$\Delta h_0=\Delta H_0/H_0$, 
one has $\Delta h_0/\Delta$[O/H]$\sim$ 
+0.124 dex$^{-1}$, i.e. an increase of $\sim$ 6\% in the LMC-based 
$H_0$ value of any SNIa calibrator whose metallicity is 0.5 
dex larger than that of the LMC. This 
is a not 
dramatic variation, nevertheless it seems interesting to 
settle whether it works to decrease or increase 
the present dispersion of $H_0$ values.    
Within this context, it is worth noticing that the well 
known disagreement between the "high" and "low" $H_0$ values 
claimed by KP and SNP studies, respectively,   
are not due entirely to the already mentioned different  
distances to SNIa calibrating 
galaxies. As discussed by Gibson et al. (2000), the 
methodology adopted by the SNP group with KP distances leads to 
$H_0\sim$ 63  km s$^{-1}$ Mpc$^{-1}$ dex$^{-1}$,    
which is 9\% higher than the SNP average value 
($H_0\sim$ 58 km s$^{-1}$ Mpc$^{-1}$ dex$^{-1}$). Moreover, the 
calibration of SNIa luminosities presents a variety 
of different approaches and the same KP distances, coupled 
with the Suntzeff et al. (1999) procedure, 
would give $H_0\sim$ 67 km s$^{-1}$ Mpc$^{-1}$ dex$^{-1}$. 
However, the absolute value of the 
Hubble constant is out the purpose of this paper. We aim only at 
determining  
the metallicity-correction to LMC-based $H_0$ values, with 
the hope of providing new elements for reducing the present uncertainty  
of $\sim$10\% to a 5\% level.

\begin{figure}
\psfig{figure=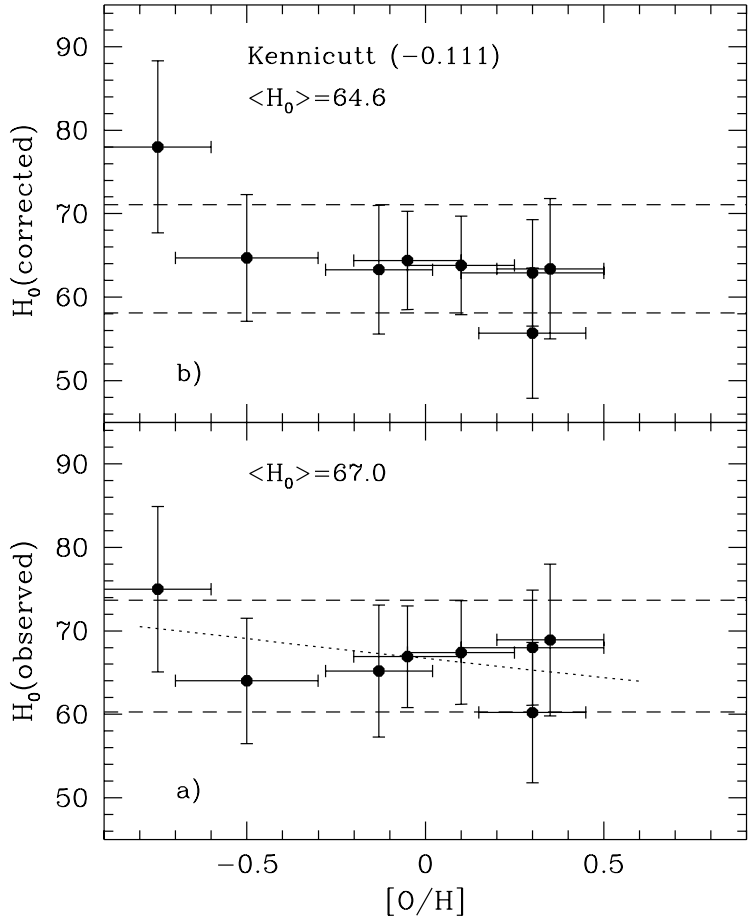,width=14truecm,angle=0} 
\caption[h]{a) The KP $H_0(V)$ values from LMC-based 
Cepheid distance to the 
eight SNIa calibrating galaxies. The average value 
is labelled. The dashed  lines depict the uncertainty of 
10\%. The dotted line depicts 
Eq. (6) in the text; 
b)  As below, but with the Kennicutt et al. (1998) metallicity 
correction to the absolute distance moduli. The resulting 
correction to $H_0$ is $\Delta h_0$/$\Delta$[O/H]$\sim-$0.111 
dex$^{-1}$ (see text).}
\end{figure}

\begin{figure}
\psfig{figure=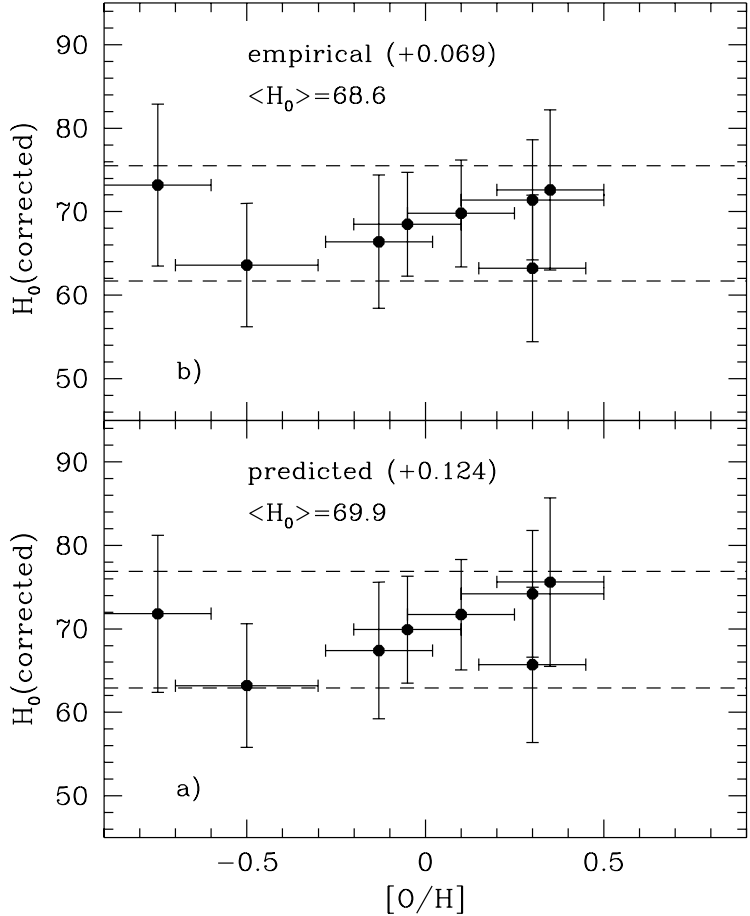,width=14truecm,angle=0} 
\caption{a) As in Fig. 5b, but with the predicted 
metallicity correction   
$\Delta h_0$/$\Delta$[O/H]$\sim$+0.124 dex$^{-1}$; 
b) As below, but with the  
metallicity 
correction 
$\Delta h_0$/$\Delta$[O/H]$\sim$+0.069 dex$^{-1}$, as suggested  
by the dotted line in  
Fig 5a.}
\end{figure}

Taking at the face value the $H_0(V)$ estimates given by 
Gibson et al. (2000, see their Table 6) 
we show in the lower panel of Fig. 5 that they agree to each other 
to within 10\% (dashed lines), but 
with a mild tendency to increase 
as the oxygen abundance of the host galaxy decreases. 
The linear regression to the points (dotted line) is
  
$$\log H_0=1.82-0.03 [O/H],\eqno(6)$$

\noindent
suggesting a 
metallicity correction as 
$\Delta h_0/\Delta$ [O/H]$\sim$ 
+0.069 dex$^{-1}$ 
which is roughly half the amount   
predicted on the basis of Eq. (3), but 
it runs towards the same direction. 
On the contrary, applying the Kennicutt et al. (1998) correction to 
the true distance moduli [Eq. (5)], results in a correction as    
$\Delta h_0$/$\Delta$[O/H]$\sim-$0.111 
dex$^{-1}$
to the LMC-based $H_0$ values. The upper panel of Fig. 5
shows that this  
would produce a  more evident correlation 
between $H_0$ and [O/H], 
with two estimates 
overcoming the discrepancy  of 
10\% from the average value (dashed lines). 
On the other hand, the lower panel of 
Fig. 6 shows that our predicted metallicity correction
$\Delta h_0$/$\Delta$[O/H]$\sim$0.124 dex$^{-1}$ 
would yield 
$H_0$ values which agree to within 10\%, 
but slightly increasing as 
the [O/H] abundance increases. Eventually, in order to 
remove any dependence of $H_0$ on the galaxy metallicity (see upper panel 
in Fig. 6), 
we suggest that
{\it at least}  
the empirical evidence in Fig. 5a should be taken into consideration, 
i.e., $\Delta h_0$/$\Delta$[O/H]$\sim$0.069 dex$^{-1}$, 
leading to a slight upward revision of   
the KP unweighted mean of  
$<H_0>$=67.0$\pm$4.3 km s$^{-1}$ Mpc$^{-1}$ 
to $<H_0>$=68.6$\pm$3.9 km s$^{-1}$ Mpc$^{-1}$. 

As said before, 
the determination of $H_0$ deals with several 
factors apart from the distance to the SNIa calibrating galaxies. 
Thus, we are \underline {not} giving the predicted 
value of the Hubble constant, but only 
the result of the predicted metallicity-correction to the KP  
$H_0$ estimates,  
still holding their adopted approach in the treatment of SNIa 
data and distance to LMC. 

\section{Concluding remarks}

HST galaxies with published Cepheid distances have been studied 
to the light of a theoretical pulsational scenario based 
on nonlinear, nonlocal and time-dependent convective models with 
three selected metallicities ($Z$=0.004, 0.008, 0.02). 

Theoretical PL and PLC relations in the $VI$ passbands 
have been used to predict Cepheid distance and 
reddening for each adopted 
metal content. The comparison of 
the predictions  
with $Z$=0.008 (the average metallicity of the LMC) 
to the LMC-calibrated 
empirical results provided by the KP group 
shows a mutual agreement, 
better than that found with SNP results. 

The predicted PL and PLC relations suggest that 
the Cepheid distance should decrease with increasing the 
adopted metallicity and that 
LMC-based distance moduli should be corrected according to  
$c=-$0.27$\Delta \log Z$ mag dex$^{-1}$, 
where $\Delta \log Z$ is 
the difference between the Cepheid metallicity and the LMC mean value 
of $Z$=0.008. 
Theoretical studies which adopt a 
linear pulsational approach, being unable to get firm predictions 
on the red edge of the instability strip, cannot disprove 
this result. 

On the observational front, earlier evidences suggest an opposite 
metallicity correction as given by 
$c$=+0.24$\Delta$[O/H] mag dex$^{-1}$ (Kennicutt et al. 1998), where 
$\Delta$[O/H] is the O/H metallicity difference between  the 
Cepheid field and the LMC mean value ([O/H]$\sim-$0.40). However, 
we show that the KP true distance modulus to galaxies members 
of clusters or groups seems to be correlated with the galaxy 
O/H metallicity in good agreement with our predictions. 

Moreover, we wish to mention that some recent  
observational clues give a further support to  
our theoretical scenario. 
They are: 
\begin{itemize}
\item The comparison between Baade Wesselink radii for Galactic 
and Magellanic Clouds Cepheids 
(Laney 1998, 1999, 2000; Feast 1999; Storm 2000) suggests that
metal-rich variables are fainter than the metal-poor ones. 
\item The purely ``geometric'' distance to 
the galaxy NGC 4258, as 
obtained from maser sources rotating around the central black hole, 
is 7.2$\pm$0.5 Mpc (Herrnstein et al. 1999), whereas the value derived 
from LMC-calibrated PL relations is 8.1$\pm$0.4 Mpc (Maoz et al. 1999).  
Since for this galaxy one has [O/H]$\sim$0.12 (Zaritsky et al. 1994), the 
Kennicutt et al. (1998) correction would 
increase the distance to  
$\sim$ 8.6 Mpc, 
whereas our predicted correction gives 
$\sim$7.6 Mpc.
\item	 A carefully selected sample of 236 Cepheids 
from the HIPPARCOS catalogue suggests that 
the slope of the Galactic PL relations could be shallower than 
that observed for LMC variables (Groenewegen \& Oudmaijer 2000), 
thus suggesting some caution in the adoption of 
universal PL relations. 
\end{itemize}

As a whole, both the predictions and the    
empirical evidences seem to  
confirm  
that the metallicity effect on the Cepheid distance scale  
is not dramatic, with a correction factor of 
$\sim-$0.14 mag to the LMC-based distance modulus of any galaxy 
whose metallicity is 0.5 dex larger than that of the 
LMC. 

The consequent 
predicted metallicity correction to the Hubble constant is   
$\Delta h_0$/$\Delta$[O/H]$\sim$+0.124 
dex$^{-1}$, which means an increase 
of $\sim$ 6.2\% in the LMC-based $H_0$ values of any 
SNIa calibrator     
whose metallicity is 0.5 dex larger than that of the
LMC. 

Taking the 
KP $H_0$ estimates at their face value, we
show that the observed correlation with the [O/H] metallicity 
of the Cepheid fields  suggests  
$\Delta h_0$/$\Delta$[O/H]$\sim$+0.069
dex$^{-1}$, which is more consistent 
with our prediction rather than the metallicity correction 
based on the Kennicutt et al. (1998) empirical results 
( $\Delta h_0$/$\Delta$[O/H]$\sim-$0.111
dex$^{-1}$.) On the basis of this, we  
eventually suggest that the 
average value provided by the KP team $<H_0>\sim$67  
km s$^{-1}$ Mpc$^{-1}$ should be increased {\it at least} up 
to $<H_0>\sim$ 69 Km s$^{-1}$ Mpc$^{-1}$.

The adoption of a metallicity correction to 
LMC-based empirical estimates of $H_0$ 
might hopefully reduce the present  
uncertainty of 10\% on the average value 
to a 5\% level. Unfortunately, the  variety of the
current methods 
of calibrating SNIa luminosities give very different 
results for the absolute value of the Hubble constant. 
As an example (see Gibson et al. 2000), the 
Suntzeff et al. (1999) methodology yields a mean 
value which is from 3.7\% to 8.3\% larger than that found with  
the Saha et al. (1997) procedure, even adopting the same distance to 
the calibrators. In our belief this is a key problem on the route 
to settle the absolute value of the Hubble constant.

\begin{acknowledgements}
We are greatly indebted to our referee for several helpful comments. 
We thank V. Castellani and G. Bono for discussion and 
critical suggestions. 
Financial support for this work was provided by the Ministero dell'Universit\`a e della Ricerca Scientifica e Tecnologica (MURST) under the scientific project
``Stellar Evolution'' (Vittorio Castellani, coordinator).
\end{acknowledgements}

\end{document}